\documentclass[final] {aipproc}
\usepackage{psfig}
\usepackage{graphicx}
\layoutstyle{6x9}
\begin{document}

   \title{V838 Mon and the new class of stars erupting 
          into cool supergiants (SECS)}

   \author{U. Munari} {
     address={Osservatorio Astronomico di Padova -- INAF, Sede di Asiago,
     I-36012 Asiago (VI), Italy}}
   \author{A.Henden} {
      address={Univ. Space Research Ass./U. S. Naval Obs.,
      P. O. Box 1149, Flagstaff AZ 86002-1149, USA}}
   \author{R.M.L.Corradi} {
      address={ING, Apartado de Correos 321,
      38700 Santa Cruz de La Palma, Canarias, Spain}}
   \author{T.Zwitter} {
      address={University of Ljubljana, Dept. of Physics, Jadranska 19, 
      1000 Ljubljana, Slovenia}}

\begin{abstract}
V838~Mon has undergone one of the most mysterious stellar outbursts on
record. The spectrum at maximum closely resembled a cool AGB star, evolving
toward cooler temperatures with time, never reaching optically thin
conditions or showing increasing ionization and a nebular stage.  The latest
spectral type recorded is M8-9. The amplitude peaked at $\Delta V$=9 mag,
with the outburst evolution being characterized by a fast rise, three maxima
over four months, and a fast decay (possibly driven by dust condensation). 
BaII, LiI and $s-$element lines were prominent in the outburst spectra.
Strong and wide (500 km/sec) P-Cyg profiles affected low ionization species,
while Balmer lines emerged to modest emission only during the central phase
of the outburst. A light-echo discovered expanding around the object
constrains its distance to 790$\pm$30 pc, providing $M_V=+4.45$ in
quiescence and $M_V=-4.35$ at optical maximum (dependent on the still
uncertain $E_{B-V}$=0.5 reddening). The visible progenitor resembles a
somewhat under-luminous F0 main sequence star, that did not show detectable
variability over the last half century.

V838~Mon together with M31-RedVar and V4332~Sgr seems to define a new class
of astronomical objects, {\sl Stars that Erupt into Cool Supergiants
(SECS)}.  They do not develop optically thin or nebular phases, and deep
P-Cyg profiles denounce large mass loss at least in the early outburst phases. 
Their progenitors are photometrically located close to the Main Sequence, away
from the post-AGB region. After the outburst, the remnants still closely
resemble the precursors (same brightness, same spectral type). Many more
similar objects could be buried among poorly studied variable stars that
have been classified as Miras or SemiRegulars on the base of a single
spectrum at maximum brightness.
\end{abstract}

\maketitle

\section{The outburst of V838 Mon}

A detailed description of the outburst of V838~Mon is given by \cite{Munari}, 
to which the reader is referred. In this note only the main features
are summarized with some updates on the late photometric and spectroscopic 
evolution of the outburst.

An updated lightcurve of the eruption of V838~Mon is presented in Figure~1.
A first maximum was reached by $+10^d$ (see abscissae scale on Figure~1)
when the continuum energy distribution was characterized by a temperature of
4150~K.  A second maximum at $+37^d$ peaked around 5200~K and a third one at
$+68^d$ reached 4600~K. Each decline from maxima was accompanied by a
monotonic cooling, with the last one taking V838~Mon to 3500~K by 
$+90^d$. From $+90^d$ to $+120^d$ the color temperature in the region of
the $V,R_c, I_c$ bands decreased to that of an M8-9 supergiant, or 2600~K.
The retracing of the $U-B$ and $B-V$ color indexes when the spectrum
developed the coolest temperatures is a real effect (cf. absolute
spectrophotometry in Figure~2 at $+119^d$), and it is normally seen in
M-type stars of the corresponding spectral types due to progressive
disappearance of TiO absorptions at the shortest wavelengths.

\begin{figure}[t]
\centering
\includegraphics[width=15.6cm]{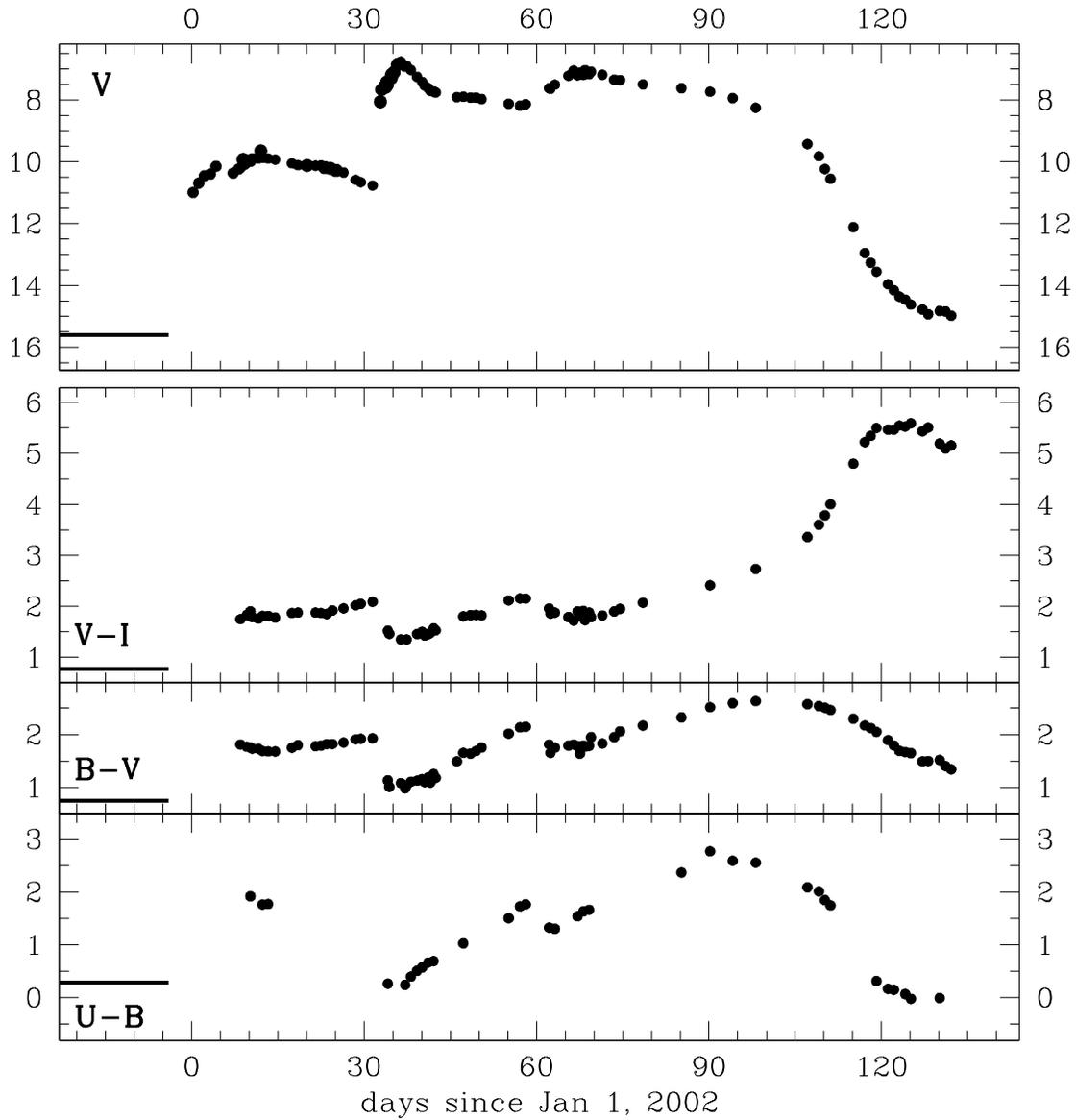}
      \caption{$V$, $B-V$ and $V-I_C$ lightcurves of the outburst of
               V838~Mon.  Dots mark NOFS data, open circles Tsukuba data.
               Crosses and open triangles are values from various
               IAUC and VSNET circulars (mainly from SAAO, D.West,
               P.Sobotka, L.Smelcer, F.Lomoz and J.Bedient). The
               solid line indicates the quiescence brightness.}
\label{lightcurve}
\end{figure}

\begin{figure}[h]
\centering
\includegraphics[width=14.8cm]{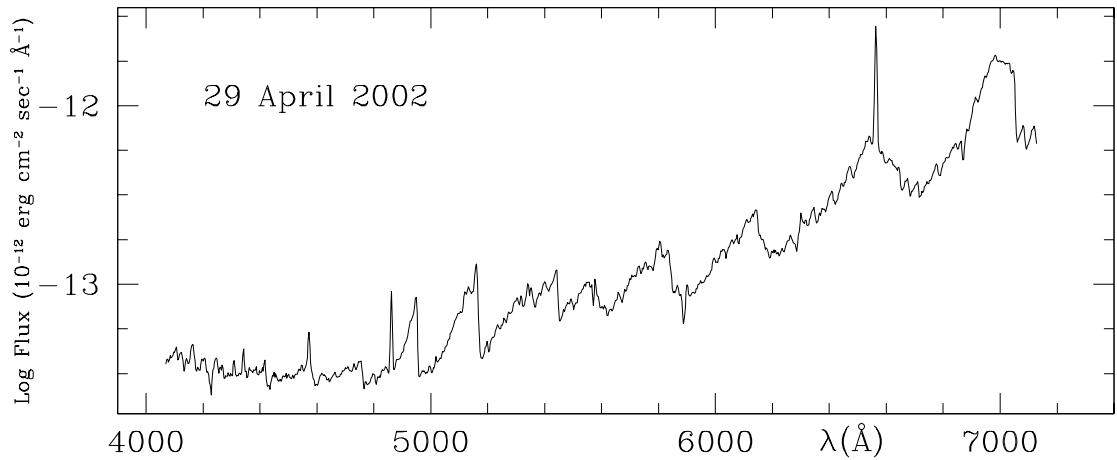}
      \caption{The spectrum of V838~Mon for April 29,
               2002 obtained with WHT 4.2m in La Palma.}
\label{Decline_spectrum}
\end{figure}

\begin{figure}[h]
\centering
\includegraphics[angle=270,width=14.8cm]{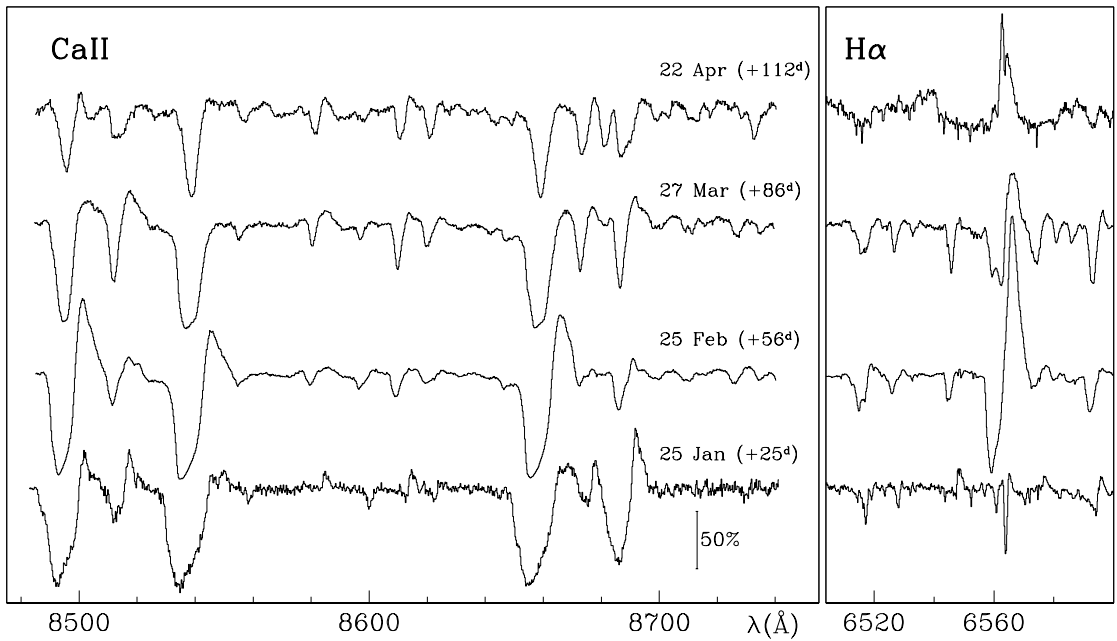}
      \caption{Small sections of sample of Asiago Echelle spectra to 
               document the evolution around the far-red Calcium triplet 
               and H$\alpha$ of the V838~Mon outburst.}
\label{CaII}
\end{figure}

\begin{figure}[h]
\centering
\includegraphics[width=14.8cm]{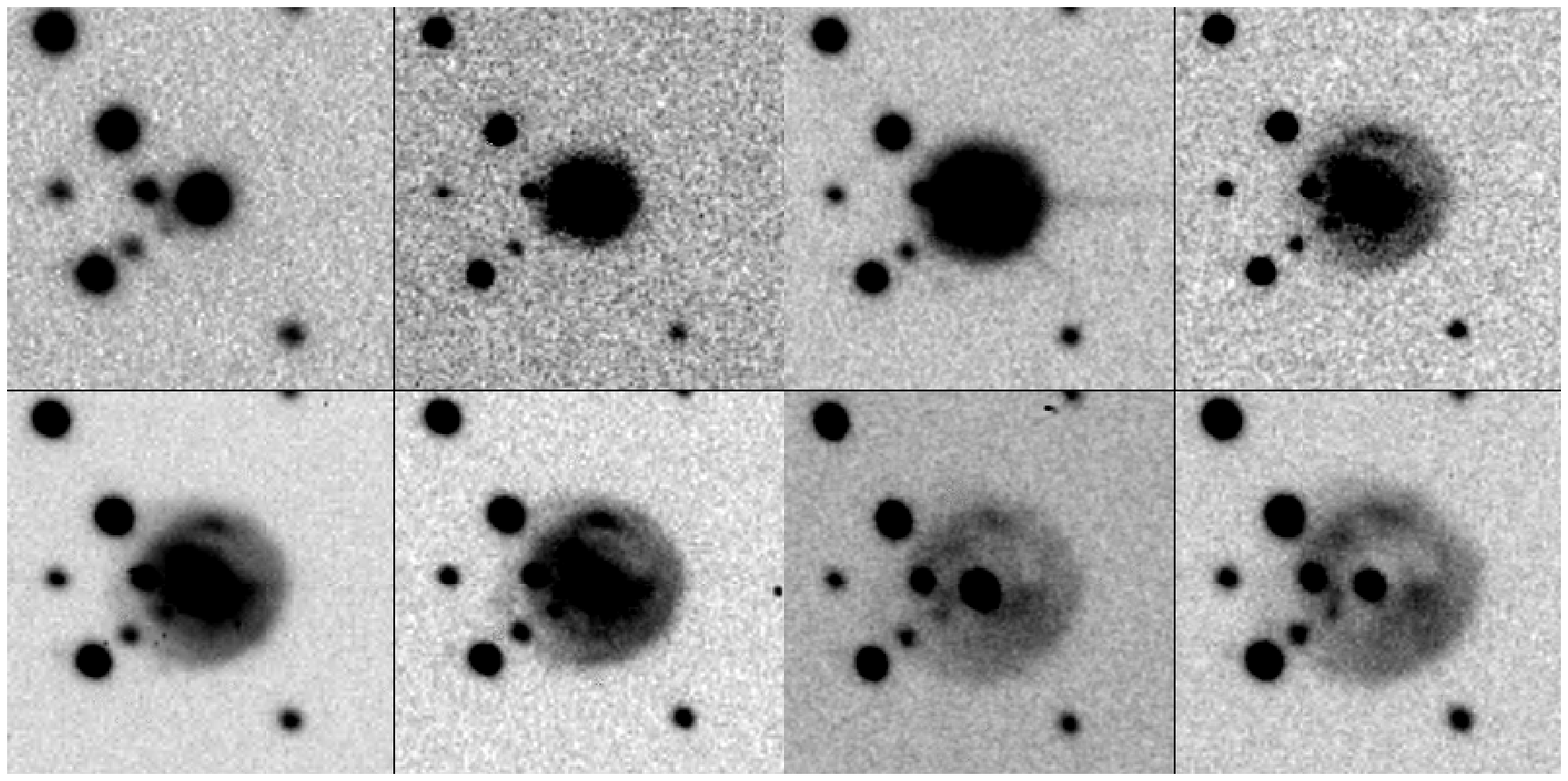}
      \caption{Expansion of the light-echo around V838~Mon, revealing a
               previously invisible ring of circumstellar material. $U$ band
               67$\times$67 arcsec images obtained with the USNO 1m
               telescope. Dates from left to right and top to bottom:
               January 13, February 27, March 10, March 27, March 31, April 4, April 20
               and April 30.}
\label{ring}
\end{figure}

The spectral evolution well followed the $V-I$ color temperature evolution.
During the first three months the spectrum closely resembled a K giant,
slowing progressing toward later spectral types and reaching K5 by
$+90^d$. In the following month the spectrum rapidly entered the M-type realm
and reached M8-9 by $+119^d$. The spectrum of V838 Mon for this date is
shown in Figure~2. The spectral evolution of V838~Mon has been exciting also
on a much finer scale, as Figure~3 indicates, where Echelle spectra around
the CaII far-red triplet and H$\alpha$ are presented for $+25^d$, $+56^d$,
$+86^d$ and $+112^d$. P~Cyg line profiles for low-excitation species have
a terminal velocity which monotonically decreased with time from the initial
value of $-$500 km/sec, with Balmer lines appearing in emission with their own P-Cyg
profiles only after the second maximum. BaII, LiI and $s-$elements are
present in the V838~Mon spectra.

In mid-February, \cite{Henden} discovered the formation
of a light-echo around V838~Mon, when the light from the second maximum
began illuminating pre-existing circumstellar material responsible for the
IRAS detection of the precursor. This light-echo was followed as it expanded
to a maximum diameter of 35 arcsec, a size that has remained essentially
constant during the following months, as Figure~4 shows. The light-echo
expansion rate of 0.44$\pm$0.017 arcsec day$^{-1}$ sets the distance of
V838~Mon to 790$\pm$30 pc for a spherical distribution of the scattering
material. The outburst light sweeping through the circumstellar material
allows us to read the recent mass loss history of the progenitor: assuming
15 km~sec$^{-1}$ velocity for its wind (typical for an AGB), the light-echo
has reached by April~1 material lost $\sim$4900 years ago. High resolution
imaging with HST by \cite{Bond} confirms the spherical symmetric dust
distribution around V838~Mon and reveals multiple circularly-symmetric
rings, along with a central void. This void was also visible on ground-based
images after V838~Mon faded.  The void is the likely reason why the
light-echo was not seen until some time after the rise to second maximum. 
The angular separation of the concentric rings in the HST images indicate a
$\sim$500 year recurrence time in the enhanced mass loss events.

Using the 790pc distance estimate and the peak brightness, we can derive
$M_V=+4.45$ for V838~Mon in quiescence and $M_V=-4.35$ at peak outburst. The
precise values depend on the exact amount of reddening, here estimated to be
$E_{B-V}$=0.5.  At galactic coordinates $l=217.80$ $b=+1.05$, the height
over the Galactic plane is just $z=$13 pc. It is relevant to note that the
progenitor of V838~Mon was not detected by H$\alpha$ emission-line surveys
in the region (these surveys discovered several faint emission line stars
close to V838~Mon), and inspection of Palomar and SERC plates as well as
results from many archival plates presented at this conference by
\cite{Barsukova} reveal absence of photometric variability in quiescence.
Both H$\alpha$ emission and variability would have supported an interactive
binary nature of the precursor.

\section{V838~Mon and the stars erupting into cool supergiants (SECS)}

In 1989 an erupting star in the Andromeda Galaxy (M31) developed a M-type
cool supergiant spectrum at maximum, with pronounced P-Cyg profiles and
Balmer lines in emission (\cite{Rich}, \cite{Mould}). The progenitor was too
faint to be identified and the event has been modeled by \cite{Iben} in
terms of a cool WD accreting at a very low rate from a companion and under
such circumstances the entire WD could experience a thermonuclear runaway.
The similarities with V838~Mon are the cool supergiant spectrum at maximum,
the emission in the Balmer lines, the variable P-Cyg profiles and the
faintness of the progenitor. However, the event in M31 peaked to $M_V =
-9.95$, much brighter that the $M_V = -4.45$ reached by V838~Mon.
Nevertheless, the light curve presented by \cite{Sharov} is quite 
similar to V838~Mon.

Another close match is Nova Sgr 1994 V4332~Sgr (Nova Sgr 1994 \#1).  As
described in \cite{Martini}, V4332~Sgr displayed a flat outburst, dramatic
increase in $(V-I)$ during its fade from maximum, an M-type spectrum at
maximum and P-Cyg profiles. The progenitor was a K star close to the Main
Sequence. The outburst duration was only 20 days, however, and perhaps this
variable was not discovered until some time after the beginning of the
outburst. Imagery during outburst as well as more recently does not show any
signs of a light-echo surrounding V4332~Sgr.

There have been several possible novae reported over the past century that
have been ascribed to Mira variability on the base of the spectrum in
outburst. Perhaps some of these are similar objects to M31~RedVar, V4332~Sgr
and V838~Mon.  The latter may represent a new class of astronomical
objects: stars erupting into cool supergiants but that never develop an
optically thin phase (increasing excitation temperature and development of a
nebular spectrum) like in classical novae. Their progenitors lies away from the
post-AGB stars and appear close to the cool main sequence on the HR diagram.
These stars, characterized by havy mass loss at least in the early outburst
phases, could be binaries even if no evidence of this has been found (in
V838~Mon the circumstellar dust envelope could be the relic of an AGB phase
of an hypotetical companion to the F0~V star seen in quiescence). We suggest
for this new class of astronomical objects the name of {\sl stars erupting
into cool supergiants} (SECS).


\begin{thebibliography}{}
\bibitem{Munari} Munari, U., Henden, A., Kiyota, S., et al., A\&ALett 389, L51 (2002)
\bibitem{Henden} Henden, A., Munari, U., Schwartz, M.B., IAUC 7859 (2002)
\bibitem{Bond} Bond, H.E., Panagia, N., Sparks, W.B., Starrfield, S.G., Wagner, R.M., IAUC 7892 (2002)
\bibitem{Barsukova} Barsukova, E.A., Borisov, N.V., Goranskij, V.P., et al., this volume (2002)
\bibitem{Rich} Rich, R.M., Mould, J., Picard, A., et al., ApJ 341, L51 (1989)
\bibitem{Mould} Mould, J., Cohen, J., Graham, J.R., et al., ApJ 353, L35 (1990)
\bibitem{Iben} Iben, I.Jr, Tutukov, A.V., ApJ 389, 369 (1992)
\bibitem{Sharov} Sharov, A.S., Sov. Astron. Lett. 19, 33 (1993)
\bibitem{Martini} Martini, P., et al., AJ 118, 1034 (1999)
\end{thebibliography}
\end{document}